\documentclass[12pt,english]{article}
\date{}
\usepackage{xcolor}
\usepackage[utf8x]{inputenc}
\usepackage{amsmath}
\usepackage{amsfonts}
\usepackage{amssymb}
\usepackage{graphicx,color}
\usepackage{babel}
\usepackage{fontenc}
\usepackage{hyperref}

\def\be{\begin{equation}}
\def\ee{\end{equation}}

\begin{document}

\title{{\bf{Holographic entanglement thermodynamics for higher dimensional charged black hole}}}

\author{
	{\bf {\normalsize Sourav Karar}$^{a,b}$\thanks{sourav.karar91@gmail.com}},
{\bf {\normalsize Debabrata Ghorai}$^{b}$ \thanks{debanuphy123@gmail.com}},
{\bf {\normalsize Sunandan Gangopadhyay}$^{b}$\thanks{sunandan.gangopadhyay@gmail.com, sunandan.gangopadhyay@bose.res.in }}\\
$^{a}$ {\normalsize Department of Physics, Government General Degree College, Muragachha 741154, Nadia, India}\\
$^{b}${\normalsize S.N. Bose National Centre for Basic Sciences, Block-JD, Sector III, Kolkata 700106, India}\\
\\[0.3cm]}

\maketitle
\begin{abstract}
In this paper, we have investigated the entanglement thermodynamics for $d$-dimensional charged $AdS$ black hole by studying the holographic entanglement entropy in different cases. We have first computed the holographic entanglement entropy in extremal and non-extremal cases in two different regimes, namely, the low temperature and high temperature limits. We then obtain the first law of entanglement thermodynamics for boundary field theory in the 
low temperature regime in $d$-dimensions. 
\end{abstract}

\newpage

\section{Introduction}
\noindent There has been an immense amount of interest in information theory recently. This is primarily due to the fact that quantum correlations, which is an important ingredient in the theory of information, play a crucial role in various branch of physics, namely, condensed matter physics, statistical mechanics as well as in quantum theories of gravity. 
It has been realized that the fundamental laws of physics can be given an information theoretic interpretation \cite{info1, info2}.
In classical information theory, information is quantified by a measure called Shannon entropy. 
The counterpart of this concept in quantum information theory is entanglement entropy (EE).
EE is a fundamental quantity in quantum information theory as it provides a measure for quantum correlation in a bipartite quantum system.
In the last few years EE has been successfully used as a probe in quantum phases 
of matter \cite{Calabrese:2004eu}-\cite{Wen:2006}.
Other areas where information theory has provided important insights are the thermodynamic derivation of Einstein's 
equations of general relativity \cite{jacob} and in resolving the black hole information loss paradox \cite{Hawking}-\cite{Maldacena:2001kr}. 
So it is realized
that in understanding the geometry of spacetime, information theory and specially EE would play a vibrant role. 

Obtaining the EE of $1+1$-dimensional conformal field theories has been an important problem in theoretical physics. The computation was first done in 
\cite{Calabrese:2004eu}-\cite{Calabrese:2009qy} by a method known as the replica trick. Interestingly, the behaviour of the EE for a $1+1$-dimensional conformal field theory (CFT) exhibited an universal logarithmic behaviour \cite{Calabrese:2004eu}. Recently, the gauge/gravity correspondence has played a key role in computing the EE of a boundary CFT holographically from its bulk gravitational dual. The insight comes from the fact that the holographic principle \cite{thooft}-\cite{Gubser:1998bc} states that the number of degrees of freedom
in a region of space is equal in number to the degrees of freedom on the boundary that surrounds the space. This principle first proposed in the context of black hole entropy, became one of the most cherished ideas in modern theoretical physics with the advent of the $AdS/CFT$ correspondence \cite{Maldacena:1997re, Aharony:1999ti}. This correspondence is the most successful realization of the holographic principle in theoretical physics. It relates the gravitational theory in $AdS$ space to the CFT that lives on the boundary of $AdS$ space. It has
evolved as a powerful theoretical input in condensed matter physics,
nuclear physics and QCD \cite{Hartnoll:2009sz}-\cite{CasalderreySolana:2011us}.

\noindent The prescription of computing the holographic entanglement entropy (HEE) was first proposed in \cite{Ryu:2006bv},\cite{Ryu:2006ef}. According to the prescription, the HEE for a subsystem $A$ in a $(d-1)$-dimensional boundary
field theory is given by
\begin{equation}
 S_A =\frac{Area (\Gamma_A)}{4 G_{(d)}} \nonumber
\end{equation}
where $\Gamma_A $ is the minimal surface area of the bulk extension (on a fixed time slice) 
whose boundary coincides with the edges of the subsystem living at the boundary and
$G_{(d)}$ is the $d$-dimensional Newton's gravitational constant. This formula is very similar to the black hole entropy
formula suggested by Bekenstein and Hawking \cite{hawk1}-\cite{hawk3}
\begin{eqnarray}
S_{BH} = \frac{Area(\sigma)}{4G_{d}} \nonumber
\end{eqnarray}
where $\sigma$ is the area of the horizon of the black hole. This striking similarity between HEE and black hole entropy inspired many to suggest that that EE is the origin of black hole entropy \cite{Bombelli:1986rw}-\cite{Eisert:2008ur}. 

Studies of the HEE of $AdS$ black holes which are dual to a field theory at finite temperature have also been carried out \cite{Cadoni:2010kla}-\cite{Fischler:2012ca}. In \cite{Chaturvedi:2016kbk}, the HEE has been computed for charged $AdS$ black hole to observe the effect of temperature and charge on the EE of a strip like subregion in the boundary field theory dual to the charged $AdS$ black hole in the bulk. An important question in this regard is whether the HEE satisfies a relation analogous to the first law of thermodynamics,
which is indeed satisfied by thermal entropy. In \cite{Bhattacharya:2012mi}, the difference in HEE of a thermally excited
$AdS$ spacetime and pure $AdS$ spacetime has been computed. This led them to conclude that the change in HEE ($\Delta S_{E}$) is proportional to the change in energy ($\Delta E$).
The proportionality constant was identified as the inverse of entanglement temperature ($T_{ent}$).
The same question was addressed in \cite{Allahbakhshi:2013rda} and
a similar relation between $\Delta S_{E}$ and $\Delta E$ was obtained. Entanglement thermodynamics has been explicitly studied in different backgrounds including non-conformal and non-relativistic backgrounds \cite{Mansoori:2015sit}-\cite{skarar}.
Entanglement thermodynamics for charged black hole in $AdS_4$ background has also been studied in \cite{Chaturvedi:2016kbk}. In this investigation, 
a relation like the first law of entanglement thermodynamics in the low temperature limit was obtained.

In this paper, we extend the analysis for a charged black hole in $d$-dimensions. We look at different temperature and charge limits.
We have explicitly calculated the EE for charged black hole ($AdS$-RN black hole) in $d$-dimensions for different temperature and charge limits. A law like the first law of entanglement thermodynamics has also been obtained in the small temperature limit. The expressions for the EE in different limits have explicit  dependence on the dimension of spacetime $d$. The analysis helps us to understand the implications of the dimension of spacetime
on information theoretic quantities.

The paper is organized as follows. We have first reviewed the $AdS$-RN black hole in arbitrary spacetime dimension in section 2. From the definition of black hole temperature, we then find the extremality condition for the $AdS$ charged black hole.
In section 3, we have investigated the HEE in small charge and large charge limit for extremal black hole. Then we have computed the HEE in the low temperature and high temperature regime for small charged non-extremal black hole as well as for large charged non-extremal black hole. 
The form of the HEE expression for small charged extremal black hole is the same as the HEE in the low temperature regime for small charged non-extremal black hole. 
It is to be noted that the bulk dual extremal black hole is considered to be the ground state.
Similarly the bulk dual non-extremal black hole is considered as the excited state. 
We then discuss the first law of entanglement thermodynamics in the small charge limit in section 4. We conclude in section 5.

\section{Charged $AdS$ black hole}

In this section, we start by writing down the charged $AdS$ black hole metric with planar horizon in $d$-dimensions. This reads
\begin{eqnarray}{\label{system}}
\label{ee2}
ds^{2} &=& - \frac{r^2}{R^2} f(r) dt^2 + \frac{R^2 dr^2}{r^2 f(r)} + \frac{r^2}{R^2} \left( dx_1^2 + dx_2^2 + ......+dx^2_{d-2}\right) \nonumber \\
f(r) &=& 1- \frac{M}{r^{d-1}} +\frac{Q^2}{r^{2(d-2)}}
\label{ee2b}
\end{eqnarray} 
where $R$ is the radius of $AdS$ spacetime, $M$ and $Q$ are the mass and charge of the black hole respectively.
We shall set $R=1$ for the rest of this paper.

The horizon of the black hole is given by $f(r)\rvert_{r=r_h}=0$. This yields
\begin{eqnarray}{\label{mass}}
M= r_h^{d-1} \left( 1+ \frac{Q^2}{r_h^{2(d-2)}}\right)
\end{eqnarray}
which relates the mass $M$ of the $AdS$-RN black hole and its charge $Q$. 
The Hawking temperature for this black hole is given by
\begin{equation}{\label{bhtemp}}
 T_{H} = \frac{r^2 f^{\prime}(r)}{4\pi}\Big\rvert_{r_{h}} 
 = \frac{(d-1)r_h}{4\pi} \left[1- \left(\frac{d-3}{d-1}\right)\frac{Q^2}{r_h^{2(d-2)}} \right]~.
\end{equation}
The lapse function $f(r)$ can now be expressed in terms of only the charge $Q$ and the radius of the horizon $r_h$ as
\begin{eqnarray}{\label{metric}}
f(r) = 1 - \left(\frac{r_h}{r}\right)^{d-1} + Q^2 \left(\frac{1}{r^{2(d-2)}}-\frac{1}{r^{d-1}r_h^{d-3}} \right)~.
\end{eqnarray}


\section{Computation of HEE}
With the basic setup in hand we are now ready to calculate the HEE of the 
$AdS$-RN black hole. We  consider an entangling region at the boundary 
in the form of a straight belt of width $l$ given by
\begin{equation}
 -\frac{l}{2} \le x_1 \le \frac{l}{2}; \quad 0 \le x_2, x_3,\cdots ,x_{d-2} \le L.
\end{equation}
According to the proposal in \cite{Ryu:2006bv, Ryu:2006ef} we have to find the minimal
codimension two hypersurface in the bulk whose boundary coincides with
the two ends of the interval $-\frac{l}{2} \le x_1 \le \frac{l}{2}$. Then 
the entanglement entropy is given by the minimal area divided by $4 G_{(d)}$, where
$G_{(d)}$ is the Newton's gravitational constant in $d$ dimensions.

The area of the hypersurface for the system (\ref{system}) is given by
\begin{equation}
 \mathcal{A} =  L^{d-3} \int_{-\frac{l}{2}}^\frac{l}{2} dx_1  
 \sqrt{r^{2(d-2)} + \frac{(r')^2}{f(r)} r^{2(d-4)}}
\end{equation}
where the surface is parametrized by $r=r(x_1)$. Using the standard procedure of minimization
we get
\begin{equation}{\label{area}}
\mathcal{A} = 2L^{d-3}
\int_{r_t}^{\infty} \frac{r^{d-4} dr}{\sqrt{f(r)\left\{1- \left(\frac{r_t}{r}\right)^{2d-4}\right\}}}
\end{equation}
with the minimal surface characterized by
\begin{equation}
 \frac{d r}{d x_1} =\sqrt{f(r)\; r^4 \left\{ \left(\frac{r^2}{r_t^2} \right)^{(d-2)} -1 \right\}}.
\end{equation}
where $r_t$ is the turning point of the extremal surface satisfying $r' \rvert_{r=r_t}=0$.
Integration of the above equation gives the length of the system to be
\begin{eqnarray}{\label{el}}
\frac{l}{2} =  \int_{r_t}^{\infty} \frac{r_t^{d-2} dr}{r^d\sqrt{f(r)\left\{1- \left(\frac{r_t}{r}\right)^{2d-4}\right\}}}
\label{ee4}
\end{eqnarray}
 

\noindent It is not difficult to see that the integral (\ref{area}) is divergent
as we reach the boundary. Therefore we have to introduce an infrared (IR) cutoff at
$r=r_b$, where $r_b$ is very large. This IR cutoff is related holographically
to the field theory counterpart which is the ultraviolet (UV) cutoff $a$ by the relation $r_b = 1/a$.
In field theory this UV cutoff is nothing but the lattice spacing. The finite part
of the entangling entropy can be used to study the high and low charge (or temperature) behavior
of the field theory which is dual to the $AdS$-RN black hole.

To compute the integrals \eqref{area} and \eqref{el}, we change the integration variable $r$
by $u= \frac{r_t}{r}$, which makes the lapse function
\begin{eqnarray}{\label{lapsu}}
f(u)= 1- \left(\frac{r_h}{r_t}\right)^{d-1} u^{d-1} - 
\frac{Q^2}{r_h^{d-3}}\left(\frac{u}{r_t}\right)^{d-1} + Q^2 \left(\frac{u}{r_t}\right)^{2(d-2)} ~.
\end{eqnarray}
The length of the subsystem along $x_1$ and the area of subsystem now read
\begin{eqnarray}
\label{elu}
l &=& \frac{2}{r_t} \int_0^1 du \frac{u^{d-2} }{\sqrt{ 1-u^{2(d-2)}}} \left( 1- \left(\frac{r_h}{r_t}\right)^{d-1} u^{d-1} - \frac{Q^2}{r_h^{d-3}}\left(\frac{u}{r_t}\right)^{d-1} + Q^2 \left(\frac{u}{r_t}\right)^{2(d-2)} \right)^{-1/2} \\
\mathcal{A} &=& 2 (L r_t)^{d-3} 
\int_0^1 du \frac{u^{-(d-2)}}{\sqrt{1-u^{2(d-2)}}}\left( 1- \left(\frac{r_h}{r_t}\right)^{d-1} u^{d-1} - \frac{Q^2}{r_h^{d-3}}\left(\frac{u}{r_t}\right)^{d-1} + Q^2 \left(\frac{u}{r_t}\right)^{2(d-2)} \right)^{-1/2}. \nonumber
\label{areau} \\
\end{eqnarray}
In the following sections we  compute the HEE for both the extremal black hole and
the black hole at finite temperature. The extremal black hole is characterized by its
zero Hawking temperature ($T_{H} = 0$). An important relation that arises in this context is the relation between the charge $Q$ of the black hole and its horizon radius $r_h$
\begin{eqnarray}{\label{extremality}}
r_h^{2(d-2)} \geq \left(\frac{d-3}{d-1}\right) Q^2 ~ .
\end{eqnarray}
In the above expression, the equality sign holds for the extremal black hole and the inequality holds for the non-extremal black hole.
It is interesting to note that the $AdS$/CFT correspondence tells that the field theory
counterpart of the $AdS$-RN black hole is in its ground state for the extremal black hole 
and in its excited state for the non-extremal black hole.

\subsection{Extremal black hole} 
Let us first calculate the HEE for the case of the extremal black hole.
The extremality condition says that the charge of the black hole has to be 
\begin{eqnarray}
Q^2 = \left( \frac{d-1}{d-3}\right) r_h^{2(d-2)} ~.
\label{ee14}
\end{eqnarray}
Using the above condition in eq.(\ref{lapsu}), we can rewrite the lapse function in terms of $u$ as
\begin{eqnarray}{\label{lapseue}}
f(u) = 1- \frac{2(d-2)}{d-3} \left(\frac{r_h u}{r_t}\right)^{d-1} + \frac{d-1}{d-3} \left(\frac{r_h u}{r_t}\right)^{2(d-2)} ~.
\end{eqnarray}
We can also express the integrals \eqref{elu},\eqref{areau} as
\begin{eqnarray}
\label{elue}
 l &=& \frac{2}{r_t} \int_0^1 du \frac{u^{d-2} }{\sqrt{ 1-u^{2(d-2)}}}
\left(1- \frac{2(d-2)}{d-3} \left(\frac{r_h u}{r_t}\right)^{d-1} + \frac{d-1}{d-3} \left(\frac{r_h u}{r_t}\right)^{2(d-2)} \right)^{-1/2}\\
\mathcal{A} &=& 2 (L r_t)^{d-3} 
\int_0^1 du \frac{u^{-(d-2)}}{\sqrt{1-u^{2(d-2)}}}
\left(1- \frac{2(d-2)}{d-3} \left(\frac{r_h u}{r_t}\right)^{d-1} + \frac{d-1}{d-3} \left(\frac{r_h u}{r_t}\right)^{2(d-2)} \right)^{-1/2}.
\label{areaue} \nonumber \\
\end{eqnarray}
It is not hard to see that these integrals cannot be evaluated analytically.
In order to evaluate the integrals analytically we have to take certain limits. In the subsequent discussion we have 
taken two extreme limits: small charge limit and large charge limit. In the following subsections
we have considered these two limits to calculate the HEE.
\subsubsection{Small charge limit}
We can see from eq.(\ref{ee14}) that if $Q$ is small, then $ r_h $ will be small. 
To be specific  we have $ l \left( \frac{d-3}{d-1} \right)^{\frac{1}{2(d-2)}} Q^{\frac{1}{d-2}} \le l r_h \ll 1$
 in the small charge limit. As the horizon radius $r_h$ is very small, 
 the turning point $r_t$ is far away from it. Therefore $\left(\frac{r_h}{r_t}\right)$ is a very small quantity.
 So we can neglect higher order terms of $\left(\frac{r_h}{r_t}\right)$. Using this approximation we can now
 Taylor expand to write
\begin{eqnarray}{\label{taylor}}
\frac{1}{\sqrt{f(u)}} \approx 1+ \frac{d-2}{d-3} \left(\frac{r_h u}{r_t}\right)^{d-1} ~.
\end{eqnarray}
Using the above equation, we can now simplify eq.(\ref{elue}) as
\begin{eqnarray}
l &\approx& \frac{2}{r_{c}} \int_0^1 \frac{u^{d-2} du}{\sqrt{ 1-u^{2(d-2)}}} \left( 1+ \frac{d-2}{d-3} \left(\frac{r_h u}{r_t}\right)^{d-1} \right) \nonumber \\
&=& \frac{2}{r_t} \left[ \int_0^1 \frac{u^{d-2} du}{\sqrt{ 1-u^{2(d-2)}}} + \frac{d-2}{d-3} \left(\frac{r_h}{r_t}\right)^{d-1} \int_0^1 \frac{u^{2d-3} du}{\sqrt{ 1-u^{2(d-2)}}} \right]~.
\label{ee18}
\end{eqnarray}
Therefore the turning point $r_t$ reads
\begin{eqnarray}
r_t = \frac{2}{l} \left[ \sqrt{\pi} \frac{\Gamma(\frac{d-1}{2(d-2)})}{\Gamma(\frac{1}{2(d-2)})} + \frac{\sqrt{\pi}}{2(d-3)}\left(\frac{r_h}{r_t}\right)^{d-1}  \frac{\Gamma(\frac{d-1}{d-2})}{\Gamma(\frac{3d-4}{2(d-2)})} \right] ~.
\label{ee19}
\end{eqnarray}
The form of the above expression suggests that we cannot solve $r_t$ exactly. Hence using the perturbative approach, we obtain 
\begin{eqnarray}
r_t = \frac{2}{l} \left[ \sqrt{\pi} \frac{\Gamma(\frac{d-1}{2(d-2)})}{\Gamma(\frac{1}{2(d-2)})} + \frac{\sqrt{\pi}}{2(d-3)}\left(\frac{l r_h}{2}\right)^{d-1} \left( \frac{\Gamma(\frac{1}{2(d-2)})}{\sqrt{\pi}\Gamma(\frac{d-1}{2(d-2)})}\right)^{d-1}  \frac{\Gamma(\frac{d-1}{d-2})}{\Gamma(\frac{3d-4}{2(d-2)})} \right]~.
\label{ee20}
\end{eqnarray} 
Now using same approximation (\ref{taylor}), the area of the extremal surface reads
\begin{eqnarray}
\mathcal{A} = 2(L r_t)^{d-3} \left[ \int_0^1 \frac{u^{-(d-2)}du}{\sqrt{1-u^{2(d-2)}}} + \frac{d-2}{d-3} \left(\frac{r_h}{r_t}\right)^{d-1} \int_0^1 \frac{u~ du}{\sqrt{ 1-u^{2(d-2)}}} \right]~.
\label{ee21}
\end{eqnarray}
It is observed that the first integral in $\mathcal{A}$ is divergent as $u\rightarrow 0$. 
To regularize the integral we introduce the UV cut-off $\frac{r_t}{r_b}$ and add a 
counter term ($\frac{-2(Lr_b)^{d-3}}{d-3}$) in order to get a finite value of the extremal area which reads
\begin{eqnarray}
\mathcal{A}^{finite} &=& 2(L r_t)^{d-3} \left[ \int_{\frac{r_t}{r_b}}^1 \frac{u^{-(d-2)}du}{\sqrt{1-u^{2(d-2)}}} + \frac{d-2}{d-3} \left(\frac{r_h}{r_t}\right)^{d-1} \int_0^1 \frac{u~ du}{\sqrt{ 1-u^{2(d-2)}}} \right] - \frac{2(Lr_b)^{d-3}}{d-3} \nonumber \\
&=& 2(L r_t)^{d-3} \left[ \frac{\sqrt{\pi}}{(d-2)} \frac{\Gamma(\frac{3-d}{2(d-2)})}{\Gamma(\frac{1}{2(d-2)})} + \frac{\sqrt{\pi}}{(d-3)}\left(\frac{r_h}{r_t}\right)^{d-1}  \frac{\Gamma(\frac{1}{d-2})}{\Gamma(\frac{d}{2(d-2)})} \right] ~.
\label{ee22}
\end{eqnarray} 
Substituting the turning point $r_t$ from eq.(\ref{ee20}) into eq.(\ref{ee22}) and keeping terms
upto $\mathcal{O} ((lr_h)^{d-1})$ and then simplifying, we obtain
\begin{eqnarray}
\mathcal{A}^{finite} &=& \left(\frac{L}{l}\right)^{d-3}  \left[  -\frac{(2\sqrt{\pi})^{d-2}}{d-3} \left(\frac{\Gamma(\frac{d-1}{d-2})}{\Gamma(\frac{1}{2(d-2)})}\right)^{d-2} + \frac{d-2}{d (d-3)} \frac{(lr_h)^{d-1}}{4\sqrt{\pi}}  \nonumber \right.\\
&& \hspace{40mm} \left. \times \left( \frac{\Gamma(\frac{1}{2(d-2)})}{\Gamma(\frac{d-1}{2(d-2)})}\right)^2 \frac{\Gamma(\frac{1}{d-2})}{\Gamma(\frac{d}{2(d-2)})}  \right] ~. 
\label{ee23}
\end{eqnarray} 
The finite holographic entanglement entropy reads ($\frac{\mathcal{A}^{finite}}{4G_{(d)}}$) from the above equation
\begin{eqnarray}
S^{finite}_{A} = S^{AdS}_{A} + S^{ext}_{A}
\label{ee24}
\end{eqnarray}
where $S^{AdS}_{A}$ is the entanglement entropy for pure $AdS$ spacetime and the extra piece comes from the extremality of black holes. The expressions for $S^{AdS}_{A}$ and $S^{ext}_{A}$ read
\begin{eqnarray}
S_A^{AdS} = - \frac{(2\sqrt{\pi})^{d-2}}{4G_N^d (d-3)} \left(\frac{L}{l}\right)^{d-3}  \left(\frac{\Gamma(\frac{d-1}{2(d-2)})}{\Gamma(\frac{1}{2(d-2)})}\right)^{d-2}
\label{ee25}
\end{eqnarray}
\begin{eqnarray}
S_A^{ext} = \frac{L^{d-3}l^2 r_h^{d-1}}{16 \sqrt{\pi} G_{(d)}} \frac{d-2}{d (d-3)}  \left(\frac{\Gamma(\frac{1}{2(d-2)})}{\Gamma(\frac{d-1}{2(d-2)})}\right)^2 \frac{\Gamma(\frac{1}{d-2})}{\Gamma(\frac{d}{2(d-2)})} ~.
\label{ee26}
\end{eqnarray}
But we know that the relation between the mass of the extremal black hole with its horizon is given by
\begin{eqnarray}
r_h^{d-1} = \frac{d-3}{2(d-2)} M^{ext} ~.
\label{ee27}
\end{eqnarray} 
Substituting this in eq.\eqref{ee26}, we obtain
\begin{eqnarray}
S_A^{ext} = k L^{d-3} l^2 M^{ext} 
\label{ee28}
\end{eqnarray}
where 
\begin{eqnarray}
 k = \frac{1}{32\;d\; G_{(d)}\sqrt{\pi}} \left( \frac{\Gamma(\frac{1}{2(d-2)})}{\Gamma(\frac{d-1}{2(d-2)})}\right)^2 \frac{\Gamma(\frac{1}{d-2})}{\Gamma(\frac{d}{2(d-2)})} ~.
\label{ee29}
\end{eqnarray}
It is reassuring to note that the above expression reduces to the result in \cite{Chaturvedi:2016kbk} in the $d=4$ limit.

\subsubsection{Large charge limit}{\label{sel}}
In this subsection we are going to compute
the HEE of an extremal $AdS$-RN black hole whose charge $Q$ is large.
By the extremality condition (\ref{extremality}), this means that the horizon radius $r_h$ is also large.
This in turn implies $r_h l\gg 1$. As the horizon radius is very large, so we can assume
that the horizon is very close to the turning point of the extremal surface ($r_h \sim r_t$).
Now looking at the area integral (\ref{areaue}), we find that the dominant contribution
to the finite part of the integral comes from $u\rightarrow 1$ limit. On the other hand, defining $u_0 =\frac{r_t}{r_h}$, we see that
$u_0 \sim 1$. Hence most of the contributuion to the finite part of the area integral comes from the
near horizon limit. We should then Taylor expand the lapse function (\ref{lapseue})
around $u_0$ to evalute the area integral. For this Taylor expansion to be valid one must show that
$u-u_0$ is small enough. As $r_t$ and $ r_b$ are very large, $u$ is very close to $u_0$ throughout the integral.
Hence we can now Taylor expand eq.(\ref{lapseue}) and neglect higher order terms to obtain
\begin{eqnarray}
f(u) &=& f(u_0) + f^{\prime}(u_0) (u-u_0) +\frac{f^{\prime\prime}(u_0)}{2!} (u-u_0)^2 + \mathcal{O}((u-u_0)^3) \nonumber \\
& = & (d-1)(d-2) \left(1- \frac{u}{u_0}\right)^2 + \mathcal{O}((u-u_0)^3) \nonumber \\
&\simeq& (d-1)(d-2) \left(1- \frac{r_h u}{r_t}\right)^2 ~.
\label{ee30}
\end{eqnarray}
Using this approximated value of $f(u)$, the length of the entangling region becomes
\begin{eqnarray}
l &=& \frac{2}{r_t \sqrt{(d-1)(d-2)}} \int_0^1 \frac{u^{d-2} du}{\sqrt{1-u^{2(d-2)}}} \frac{1}{(1- \frac{r_h}{r_t} u)} ~.
\end{eqnarray}
To simplify this integral we make a binomial expansion and  the length integral now takes the form
\begin{eqnarray}
 \frac{lr_t}{2}&=&  \frac{1}{ \sqrt{(d-1)(d-2)}} \sum_{n=0}^{\infty} \left(\frac{r_h}{r_t}\right)^n \int_0^1 \frac{u^{n+d-2} du}{\sqrt{1-u^{2(d-2)}}} \nonumber \\
&=& \frac{1}{2(d-2)^{3/2}} \sqrt{\frac{\pi}{d-1}}  \sum_{n=0}^{\infty}  \frac{\Gamma(\frac{n+d-1}{2(d-2)})}{\Gamma(\frac{n+2d-3}{2(d-2)})} \left(\frac{r_h}{r_t}\right)^n ~.
\label{ee31}
\end{eqnarray}
For large value of $n$, this expression is divergent. Using gamma function properties and Stirling formula, one can check that for large value of $n$, the above summation goes 
as $\frac{\sqrt{2(d-2)}}{\sqrt{n}} \left(\frac{r_h}{r_t}\right)^n $. To get a finite value we isolate the divergent terms to  obtain
\begin{eqnarray}
lr_t &=& \frac{1}{(d-2)^{3/2}}  \sqrt{\frac{\pi}{d-1}}  \frac{\Gamma(\frac{d-1}{2(d-2)})}{\Gamma(\frac{2d-3}{2(d-2)})} + \sqrt{\frac{\pi}{(d-1)(d-2)}} \sum_{n=1}^{\infty} \left(  \frac{\Gamma(\frac{n+d-1}{2(d-2)})}{(d-2)\Gamma(\frac{n+2d-3}{2(d-2)})} - \sqrt{\frac{2}{(d-2)n}} \right) \left(\frac{r_h}{r_t}\right)^n \nonumber \\
&& +  \frac{1}{(d-2)} \sqrt{\frac{2\pi}{d-1}}  Li_{\frac{1}{2}}\left(\frac{r_h}{r_t}\right)
\label{ee32}
\end{eqnarray}
where 
\begin{eqnarray}
Li_{\frac{1}{2}} (\frac{r_h}{r_t}) = \sum_{n=1}^{\infty} \frac{1}{\sqrt{n}} \left(\frac{r_h}{r_t}\right)^n
\label{ee33}
\end{eqnarray}
is polylogarithmic function. As the horizon radius $r_h$ is very close to 
the turning point $r_t$ of the extremal surface, we can assume that $r_t = r_h (1+ \epsilon)$ 
where $\epsilon$ is a very small positive number \cite{Hubeny:2012ry}. Substituting it in eq.(\ref{ee32}), we get 
\begin{eqnarray}
lr_h = k_1 + \sqrt{\frac{2}{d-1}} \left(\frac{\pi}{d-2}\right)\frac{1}{\sqrt{\epsilon}} + \mathcal{O}(\epsilon)
\label{ee34}
\end{eqnarray}
where 
\begin{eqnarray}
k_1 &=& \sqrt{\frac{\pi}{d-1}} \frac{1}{(d-2)^{3/2}} \frac{\Gamma(\frac{d-1}{2(d-2)})}{\Gamma(\frac{2d-3}{2(d-2)})}  + \sqrt{\frac{2\pi}{d-1}} \frac{1}{(d-2)} \zeta (\frac{1}{2})\nonumber \\ 
&& + \sqrt{\frac{\pi}{(d-1)(d-2)}} \sum_{n=1}^{\infty} \left(  \frac{\Gamma(\frac{n+d-1}{2(d-2)})}{(d-2)\Gamma(\frac{n+2d-3}{2(d-2)})}  -\sqrt{\frac{2}{(d-2)n}} \right).
\end{eqnarray}
We can also invert eq.(\ref{ee34}) to obtain
\begin{eqnarray}
\epsilon \approx \frac{2\pi^2}{(d-1)(d-2)^2} \frac{1}{(lr_h - k_1)^2} ~.
\end{eqnarray}
The $r_h$ appearing in the above equation can be replaced by eq.(\ref{ee14}) to get $\epsilon$ in terms of the black hole charge $Q$.

\noindent Now the extremal surface area (\ref{areaue}) reads
\begin{equation}
 \mathcal{A} = \frac{2 (L r_t)^{d-3}}{\sqrt{(d-1)(d-2)}} 
\int_0^1 du \frac{u^{-(d-2)}}{\sqrt{1-u^{2(d-2)}}}\frac{1}{(1- \frac{r_h}{r_t} u)} ~.
\end{equation}
Again by using the binomial expansion we get
\begin{eqnarray}{\label{areasum1}}
\mathcal{A} = \frac{2(Lr_t)^{d-3}}{\sqrt{(d-1)(d-2)}} \sum_{n=0}^{\infty} \left(\frac{r_h}{r_t}\right)^n \int^{1}_{0} \frac{u^{n-d+2}du}{\sqrt{1-u^{2(d-2)}}} ~.
\end{eqnarray}
The above equation is divergent for the terms corresponding to $n<(d-2)$. Let us regularize it from $n=0$ to $n=d-3$.
To regularize the divergent terms we have to introduce IR cut-off $r_b$ in the integrals. Let us start with the integral
corresponding to $n=0$:
\begin{eqnarray}
 \mathcal{A}_0^{finite} &=& \frac{2(Lr_t)^{d-3}}{\sqrt{(d-1)(d-2)}}
		  \int_{\frac{r_t}{r_b}}^{1} du \frac{1}{u^{d-2}\sqrt{1-u^{2(d-2)}}} - \frac{2(L r_b)^{d-3}}{\sqrt{(d-1)(d-2)}} \nonumber \\
	&=& -\frac{2\sqrt{\pi} (Lr_t)^{d-3}}{(d-3)\sqrt{(d-1)(d-2)}} \frac{\Gamma \left( \frac{d-1}{2(d-2)}  \right)}{\Gamma \left(\frac{1}{2(d-2)}   \right)}	~.  
\end{eqnarray}
The term corresponding to $n=1$ is given by
\begin{eqnarray}
 \mathcal{A}_1^{finite} &=& \frac{2(Lr_t)^{d-3}}{\sqrt{(d-1)(d-2)}} \left(\frac{r_h}{r_t}\right) \int^{1}_{0} du  \frac{u^{-(d-3)}}{\sqrt{1-u^{2(d-2)}}} \nonumber \\
 &=& \frac{2 r_h L^{d-3} r_t^{d-4}}{\sqrt{(d-1)(d-2)}}
 \left[\int_{\frac{r_t}{r_b}}^1 du \frac{1}{u^{d-3}} + 
 \sum_{k=1}^{\infty}\frac{\Gamma (k+\frac{1}{2})}{\sqrt{\pi} \Gamma (k+1)} \int_0^1 du u^{3-d +2k(d-2)}\right].
\end{eqnarray}
In the above equation, we have separated the first term as it is divergent and we have regularized it.
Since $r_t$ and $r_b$ are both large ($r_t \sim r_b$) hence the first term do not contribute. The finite value of the integral is
\begin{equation}
  \mathcal{A}_1^{finite} =\frac{2 r_h L^{d-3} r_t^{d-4}}{\sqrt{(d-1)(d-2)}}
  \left[\frac{1}{d-4} + \frac{\sqrt{\pi}}{2(d-2)}\frac{\Gamma (\frac{4-d}{2(d-2)})}{\Gamma (\frac{2}{2(d-2)})} \right] ~.
\end{equation}
In general the expression for the regularized terms are
\begin{equation}
 \mathcal{A}_m^{finite} = \frac{2(Lr_t)^{d-3}}{\sqrt{(d-1)(d-2)}} \left(\frac{r_h}{r_t}\right)^m 
	\left[\frac{1}{d-m-3}+ \frac{\sqrt{\pi}}{2(d-2)}\frac{\Gamma (\frac{m-d+3}{2(d-2)})}{\Gamma (\frac{m+1}{2(d-2)})} \right]
\end{equation}
for $m=1,2,\dots,(d-4)$. Let us now look at the $n=d-3$ term:
\begin{eqnarray}
  \mathcal{A}_{d-3} &=& \frac{2 (Lr_h)^{d-3}}{\sqrt{(d-1)(d-2)}} \int_0^1 du \frac{1}{u \sqrt{1-u^{2(d-2)}}} \nonumber \\
  &=& \frac{2 (Lr_h)^{d-3}}{\sqrt{(d-1)(d-2)}} \left[\int_{\frac{r_t}{r_b}}^{1} du\; \frac{1}{u}+
   \sum_{k=1}^{\infty}\frac{\Gamma (k+\frac{1}{2})}{\sqrt{\pi} \Gamma (k+1)} \int_0^1 du\; u^{2k(d-2)-1}\right] \nonumber \\
   &=& \frac{2 (Lr_h)^{d-3}}{\sqrt{(d-1)(d-2)}}\left[ -\log (\frac{r_t}{r_b}) +\frac{\log 4}{2(d-2)}\right]\nonumber\\
   & \approx & \frac{2 (Lr_h)^{d-3}}{\sqrt{(d-1)(d-2)}} \frac{\log 4}{2(d-2)} ~.
\end{eqnarray}

The remaining terms in eq.\eqref{areasum1} corresponding to $n \geq (d-2)$ are given as
\begin{eqnarray}
\mathcal{A}_{n\geq (d-2)} &=& \frac{2(Lr_t)^{d-3}}{\sqrt{(d-1)(d-2)}} \sum_{n=(d-2)}^{\infty} \left(\frac{r_h}{r_t}\right)^n \int_{0}^{1} \frac{u^{n-d+2}du}{\sqrt{1-u^{2(d-2)}}} \nonumber \\
&=& \frac{2(Lr_t)^{d-3}}{\sqrt{(d-1)(d-2)}} \sum_{n=(d-2)}^{\infty} \frac{\sqrt{\pi}}{2(d-2)} \frac{\Gamma(\frac{n-d+3}{2(d-2)})}{\Gamma(\frac{n+1}{2(d-2)})} \left(\frac{r_h}{r_t}\right)^n ~.
\end{eqnarray}
Therefore the above contribution diverges as $r_t$ approaches to $r_h$, as for large $n$ the
factor inside the summation goes as $\sim \frac{1}{\sqrt{n}} \left(\frac{r_h}{r_t}\right)^n$.
To remove the divergence we use the identity $\Gamma (n+1) =n \Gamma (n)$ to rewrite it as
\begin{equation}
 \mathcal{A}_{n\geq (d-2)}=\frac{(Lr_t)^{d-3}\sqrt{\pi}}{\sqrt{(d-1)(d-2)}}  \sum_{n=(d-2)}^{\infty}  \left\{ \frac{1}{(d-2)}+ \frac{1}{(n-d+3)} \right\} \frac{\Gamma(\frac{n+d-1}{2(d-2)})}{\Gamma(\frac{n+2d-3}{2(d-2)})} \left(\frac{r_h}{r_t}\right)^n ~.
\end{equation}
Now for large value of $n$, the second term goes as $\frac{\sqrt{2(d-2)}}{n\sqrt{n}} \left(\frac{r_h}{r_t}\right)^n$. 
Therefore it is convergent. Using eq.(\ref{ee31}) in the above equation we obtain
\begin{eqnarray}{\label{3.44}}
\mathcal{A}_{n\geq (d-2)}^{finite} &=& \frac{(Lr_t)^{d-3}\sqrt{\pi}}{\sqrt{(d-1)(d-2)}} \left[ \frac{\sqrt{(d-2)(d-1)}}{\sqrt{\pi}} lr_t -\sum_{m=0}^{d-3} \frac{\Gamma(\frac{m+d-1}{2(d-2)})}{(d-2)\Gamma(\frac{m+2d-3}{2(d-2)})} \left(\frac{r_h}{r_t}\right)^m \nonumber \right. \\
 && \left. + \sum_{n=d-2}^{\infty} \ \frac{1}{(n-d+3)}\frac{\Gamma(\frac{n+d-1}{2(d-2)})}{\Gamma(\frac{n+2d-3}{2(d-2)})} \left(\frac{r_h}{r_t}\right)^n  \right] ~.
\end{eqnarray}
The leading contribution in $\mathcal{A}_{n\geq (d-2)}^{finite}$ comes from the limit $r_t =r_h$. The second series
in eq.(\ref{3.44}) is convergent at leading order. If we want to find the subleading order contributions,
we have have to put $r_t=r_h(1+\epsilon)$ and expand it binomially. Now the second series in eq.(\ref{3.44}) may not be
convergent at the subleading order. Therefore we isolate the divergent terms of the series up to $\mathcal{O}(\epsilon)$ 
and rewrite eq.(\ref{3.44}) as
\begin{eqnarray}
\mathcal{A}_{n\geq (d-2)}^{finite} &=& \frac{(Lr_t)^{d-3}\sqrt{\pi}}{\sqrt{(d-1)(d-2)}} \left[ \frac{\sqrt{(d-2)(d-1)}}{\sqrt{\pi}} lr_t -\sum_{m=0}^{d-3} \frac{\Gamma(\frac{m+d-1}{2(d-2)})}{(d-2)\Gamma(\frac{m+2d-3}{2(d-2)})} \left(\frac{r_h}{r_t}\right)^m \nonumber \right. \\
&& \left. + \sum_{n=d-2}^{\infty} \left( \frac{1}{n-d+3}\frac{\Gamma(\frac{n+d-1}{2(d-2)})}{\Gamma(\frac{n+2d-3}{2(d-2)})} -  \frac{\sqrt{2(d-2)}}{n\sqrt{n}}\right) \left(\frac{r_h}{r_t}\right)^n + \sum_{n=d-2}^{\infty} \frac{\sqrt{2(d-2)}}{n\sqrt{n}} \left(\frac{r_h}{r_t}\right)^n  \right]. \nonumber \\
\end{eqnarray}
Now we can write 
\begin{eqnarray}
\sum_{n=d-2}^{\infty} \frac{\sqrt{2(d-2)}}{n\sqrt{n}} \left(\frac{r_h}{r_t}\right)^n = \sqrt{2(d-2)} \left[ Li_{\frac{3}{2}} \left[\frac{r_h}{r_t}\right]  - \sum_{m=1}^{d-3} \frac{1}{m\sqrt{m}} \left(\frac{r_h}{r_t}\right)^m \right].
\end{eqnarray}
We can now put all the results together in eq.(\ref{areasum1}) to write the total area of the extremal surface as
\begin{eqnarray}
 \mathcal{A}^{finite} &=& \frac{(Lr_t)^{d-3}}{\sqrt{(d-1)(d-2)}} \left[   \sqrt{(d-2)(d-1)} lr_t - \frac{(d-2)\sqrt{\pi}}{d-3} \frac{\Gamma (\frac{d-1}{2(d-2)})}{\Gamma (\frac{1}{2(d-2)})} \right.\nonumber \\
&& +\sum_{n=1}^{d-4}\left(\frac{1}{d-n-3}+ \frac{\sqrt{\pi}}{2(d-2)}\frac{\Gamma (\frac{n-d+3}{2(d-2)})}{\Gamma (\frac{n+1}{2(d-2)})}- \frac{\sqrt{\pi}}{2(d-2)} \frac{\Gamma (\frac{n+d-1}{2(d-2)})}{\Gamma (\frac{n+2d-3}{2(d-2)})} \right)\left(\frac{r_h}{r_t}\right)^n \nonumber\\
&&+ \left(\frac{\log 4}{(d-2)} -\frac{\sqrt{\pi}}{(d-2)\Gamma (3/2)}\right) \left(\frac{r_h}{r_t}\right)^{d-3} \nonumber \\
&&+ \sqrt{\pi}\sum_{n=d-2}^{\infty} \left( \frac{1}{n-d+3}\frac{\Gamma(\frac{n+d-1}{2(d-2)})}{\Gamma(\frac{n+2d-3}{2(d-2)})}-  \frac{\sqrt{2(d-2)}}{n\sqrt{n}}\right)\left(\frac{r_h}{r_t}\right)^n \nonumber\\
&& \left. \sqrt{2\pi(d-2)} \left( Li_{\frac{3}{2}} \left[\frac{r_h}{r_t}\right] - \sum_{n=1}^{d-3} \frac{1}{n\sqrt{n}} \left(\frac{r_h}{r_t}\right)^n \right)\right] ~.
\end{eqnarray}
Now we substitute $r_t = r_h(1+\epsilon)$ in the above expression. After simplification we finally obtain
\begin{eqnarray}
\mathcal{A}^{finite} = L^{d-3} l r_h^{d-2} + (Lr_h)^{d-3} \left( K_{1} + K_{2} \sqrt{\epsilon} +K_3 \epsilon\right) + \mathcal{O}(\epsilon^{3/2}) 
\end{eqnarray} 
where 
\begin{eqnarray}
K_{1} &=& \frac{1}{\sqrt{(d-1)(d-2)}} \left[ -2\sqrt{\pi}\frac{d-2}{d-3} \frac{\Gamma(\frac{d-1}{2(d-2)})}{\Gamma(\frac{1}{2(d-2)})}+ \frac{\log 4}{(d-2)} -\frac{\sqrt{\pi}}{(d-2)\Gamma (3/2)}  \right. \nonumber \\   
&& \left. +\sqrt{2\pi (d-2)}\left[ \zeta\left(\frac{3}{2}\right) - \sum_{n=1}^{d-3}\frac{1}{n\sqrt{n}} \right] +\sqrt{\pi} \sum_{n=d-2}^{\infty} \left[\frac{1}{n-d+3} \frac{\Gamma(\frac{n+d-1}{2(d-2)})}{\Gamma(\frac{n+2d-3}{2(d-2)})} -\frac{\sqrt{2(d-2)}}{n\sqrt{n}} \right]  \right. \nonumber \\
&& \left.  + \sum_{n=1}^{d-3} \left(\frac{1}{d-n-3} +\frac{d-2}{n+1}\frac{\Gamma (\frac{n-d+3}{2(d-2)})}{\Gamma (\frac{n+1}{2(d-2)})}\right) \right] \nonumber \\
K_{2} &=& -\frac{2\sqrt{2}\pi}{\sqrt{d-1}} \nonumber \\
K_3 &=& \frac{2}{\sqrt{(d-1)(d-2)}} \left[1+\frac{d}{d-1}\frac{\Gamma(\frac{-1}{2(d-2)})}{\Gamma(\frac{d-3}{2(d-2)})}
+ \sqrt{\frac{\pi (d-2)}{2}}\left( (d-1)\xi(\frac{3}{2})-\xi(\frac{1}{2})\right)\right] ~.
\end{eqnarray}
Therefore the renormalized holographic entanglement entropy in the large charge regime for the extremal black hole is given by 
\begin{eqnarray}
S_{A}^{finite} = L^{d-3}l S_{BH}^{ext} + \frac{(Lr_h)^{d-3}}{4G_N^d} \left( K_{1} + K_{2} \sqrt{\epsilon}+ K_3 \epsilon\right)+\mathcal{O}(\epsilon^{3/2}) 
\end{eqnarray} 
where $S_{BH}^{ext} = \frac{r_h^{d-2}}{4G_N^d}~$.

\subsection{Non-extremal black hole} 
\subsubsection{ Small Charge limit}
\textbf{Low temperature limit} \\
In this subsection we shall compute the HEE for a subsystem in the boundary theory
whose bulk dual is the non-extremal black hole which has small charge $Q$. We further assume
that the Hawking temperature $T_H$ of the black hole is also small. The non-extremality condition implies
$Q^2 \neq \frac{d-1}{d-3} r_h^{2(d-2)}$. As the temperature of the black hole $T_H$ is small, eq.(\ref{bhtemp}) implies that $r_h$ has to be small. Again eq.(\ref{extremality}) in small charge limit also suggest the same criterion. Thus in small charge and temperature limit, the horizon radius $r_h$ of the black hole is very small, so we can make the assumption $\frac{r_h}{r_t} \ll 1$.
Now we replace the charge $Q$ by the quantity $\alpha= \frac{Q}{r_h^{d-2}}$ in eq.(\ref{lapsu}) which makes the lapse function to be
\begin{equation}
 f(u) = 1 - \left\{ 1+ \alpha^2 - \left(\frac{r_h u}{r_t}\right)^{d-3} \right\} \left(\frac{r_h u}{r_t}\right)^{d-1} ~.
\end{equation}
To find  the values of the expressions \eqref{elu} and \eqref{areau}, we have to 
use an approximated expression for $\frac{1}{\sqrt{f(u)}}$. As we have already made the assumption of 
low temperature, eq.(\ref{bhtemp}) implies that $\frac{Q}{r_h^{d-2}}=\alpha \sim 1$. 
We can now Taylor expand $\frac{1}{\sqrt{f(u)}}$ around $\frac{r_h}{r_t} \sim 0$ and neglect higher order terms
to obtain
\begin{equation}
 \frac{1}{\sqrt{f(u)}}=  1+ \frac{1+\alpha^2}{2} \left(\frac{r_h u}{r_t}\right)^{d-1} + \mathcal{O}(\left(\frac{r_h u}{r_t}\right)^{2(d-2)}) ~.
\end{equation}
Using this in eq.(\ref{elu}) we obtain the length of subsystem to be
\begin{eqnarray}
l &=& \frac{2}{r_t} \left[ \int_{0}^{1} \frac{u^{d-2}du}{\sqrt{1-u^{2(d-2)}}} + \frac{1+\alpha^2}{2} \left(\frac{r_h}{r_t}\right)^{d-1} \int_{0}^{1} \frac{u^{2d-3}du}{\sqrt{1-u^{2(d-2)}}} \right] \nonumber \\
\Rightarrow r_t &=& \frac{2}{l} \left[\sqrt{\pi} \frac{\Gamma(\frac{d-1}{2(d-2)})}{\Gamma(\frac{1}{2(d-2)})} + \frac{1+\alpha^2}{2} \left(\frac{r_h}{r_t}\right)^{d-1} \frac{\sqrt{\pi}}{2(d-2)^2} \frac{\Gamma(\frac{1}{d-2})}{\Gamma(\frac{3d-4}{2(d-2)})} \right] ~.
\label{ee51}
\end{eqnarray}
To get the solution of the turning point $r_t$ of the extremal surface in terms
of the length ($l$) of the subsystem, we use perturbative technique to get 
\begin{eqnarray}
r_t = \frac{2\sqrt{\pi}}{l} \left[\frac{\Gamma(\frac{d-1}{2(d-2)})}{\Gamma(\frac{1}{2(d-2)})} + \frac{1+\alpha^2}{(d-2)^2} \frac{(lr_h)^{d-1}}{2^{d+1}} \left(\frac{\Gamma(\frac{1}{2(d-2)})}{\sqrt{\pi}\Gamma(\frac{d-1}{2(d-2)})}\right)^{d-1} \frac{\Gamma(\frac{1}{d-2})}{\Gamma(\frac{3d-4}{2(d-2)})} \right] ~.
\label{ee53}
\end{eqnarray}
Now the extremal surface area reads
\begin{eqnarray}
\mathcal{A} = 2 (Lr_t)^{d-3} \left[ \int_{0}^{1} \frac{u^{-d+2}du}{\sqrt{1-u^{2(d-2)}}} + \frac{1+\alpha^2}{2} \left(\frac{r_h}{r_t}\right)^{d-1} \int_{0}^{1} \frac{u du}{\sqrt{1-u^{2(d-2)}}} \right] ~.
\end{eqnarray}
The first integral of the above expression is divergent. This can be regularized by introducing UV cut-off $\frac{1}{r_b}$.
So, the finite part of the extremal surface area is
\begin{eqnarray}
\mathcal{A}^{finite} &=& 2 (Lr_t)^{d-3} \left[ \int^{1}_{\frac{r_t}{r_b}} \frac{u^{-d+2}du}{\sqrt{1-u^{2(d-2)}}} + \frac{1+\alpha^2}{2} \left(\frac{r_h}{r_t}\right)^{d-1} \frac{\sqrt{\pi}}{2(d-2)} \frac{\Gamma(\frac{1}{d-2})}{\Gamma(\frac{d}{2(d-2)})} \right] - \frac{2(lr_b)^{d-3}}{d-3} \nonumber \\
&=& \frac{(Lr_t)^{d-3}\sqrt{\pi}}{d-2} \left[ \frac{\Gamma(\frac{3-d}{2(d-2)})}{\Gamma(\frac{1}{2(d-2)})} + \frac{1+\alpha^2}{2} \left(\frac{r_h}{r_t}\right)^{d-1}  \frac{\Gamma(\frac{1}{d-2})}{\Gamma(\frac{d}{2(d-2)})} \right] ~.
\end{eqnarray} 
Now we shall substitute eq.(\ref{ee53}) in the above equation to express the extremal 
surface area in terms of the subsystem length $l$. This yields
\begin{eqnarray}
\mathcal{A}^{finite} &=& \left(\frac{L}{l}\right)^{d-3} \left[ -\frac{(2\sqrt{\pi})^{d-2}}{d-3} \left(\frac{\Gamma(\frac{d-1}{d-2})}{\Gamma(\frac{1}{2(d-2)})}\right)^{d-2} + \frac{1+\alpha^2}{8\sqrt{\pi}d} (lr_h)^{d-1} \right. \nonumber\\
&& \hspace{40mm} \left. \times \left( \frac{\Gamma(\frac{1}{2(d-2)})}{\Gamma(\frac{d-1}{2(d-2)})}\right)^2 \frac{\Gamma(\frac{1}{d-2})}{\Gamma(\frac{d}{2(d-2)})} \right] ~.
\end{eqnarray}
From the definition of the entanglement entropy we therefore obtain
\begin{eqnarray}
S_{A}^{finite} = \frac{\mathcal{A}^{finite}}{4 G_N^d} = S_{A}^{AdS} + S_{A}^{non-ext}
\end{eqnarray}
where 
\begin{eqnarray}
S_A^{AdS} &=& - \frac{(2\sqrt{\pi})^{d-2}}{4 G_N^d (d-3)} \left(\frac{L}{l}\right)^{d-3} \left( \frac{\Gamma(\frac{d-1}{2(d-2)})}{\Gamma(\frac{1}{2(d-2)})}\right)^{d-2} \\
S_{A}^{non-ext} &=& \frac{(1+\alpha^2)L^{d-3}\;l^2\;r_h^{d-1}}{32 \;d \sqrt{\pi}  G_{(d)}} \left( \frac{\Gamma(\frac{1}{2(d-2)})}{\Gamma(\frac{d-1}{2(d-2)})}\right)^2 \frac{\Gamma(\frac{1}{d-2})}{\Gamma(\frac{d}{2(d-2)})} ~.
\label{ee56}
\end{eqnarray}
These expressions reproduces the results in \cite{Chaturvedi:2016kbk} in $d=4$ limit. We can now use the mass of non-extremal black hole which is $M = r_h^{d-1} (1+\alpha^2)$ to express
$ S_{A}^{non-ext}$ in a more convenient way as
\begin{equation}
S_{A}^{non-ext} = k L^{d-3} l^2M^{non-ext}
\end{equation}
where 
\begin{equation}
 k = \frac{1}{32\;d\; G_{(d)}\sqrt{\pi}} \left( \frac{\Gamma(\frac{1}{2(d-2)})}{\Gamma(\frac{d-1}{2(d-2)})}\right)^2 \frac{\Gamma(\frac{1}{d-2})}{\Gamma(\frac{d}{2(d-2)})} 
\end{equation}
which is same as eq.(\ref{ee29}).\\

\noindent \textbf{High temperature limit} \\
In this subsection we will investigate the behaviour of HEE for a subsystem in
the boundary whose bulk dual is a non-extremal black hole with small charge but
a high Hawking temperature. In the high temperature limit the expression for temperature 
of black hole \eqref{bhtemp} and the non-extremality condition \eqref{extremality} together suggest 
that the horizon radius $r_h$ has to be large. Therefore one may easily see that $\frac{Q^2}{r_h^{2(d-2)}} \ll 1 $.
We now define a new the quantity $\delta^2 = \frac{(d-3)Q^2}{(d-1)r_h^{2(d-2)}} \ll 1$ to express lapse function as
\begin{eqnarray}
f(u) &=& 1- \left(\frac{r_h u}{r_t}\right)^{d-1} -\frac{d-1}{d-3} \left(\frac{r_h u}{r_t}\right)^{d-1} \delta^2 \left(1- \left(\frac{r_h u}{r_t}\right)^{d-3} \right) \\
\Rightarrow \frac{1}{\sqrt{f(u)}} &\approx & \frac{1}{\sqrt{1- \left(\frac{r_h u}{r_t}\right)^{d-1}}} \left[1+ \frac{(d-1)\delta^2}{2(d-3)} \left(\frac{r_h u}{r_t}\right)^{d-1} \frac{1-\left(\frac{r_h u}{r_t}\right)^{d-3}}{1- \left(\frac{r_h u}{r_t}\right)^{d-1}} \right] ~.
\end{eqnarray} 
In the last line we have made a Taylor expansion of the function around $\delta =0$ and neglected the higher order terms.
We can use this expression of lapse function in eq.(\ref{elu}) to find the subsystem length to be
\begin{eqnarray}{\label{3.65}}
l = \frac{2}{r_t} &&\left[\int_0^1   \frac{u^{d-2}du}{\sqrt{1-u^{2(d-2)}}} \frac{1}{\sqrt{1-\left(\frac{r_h u}{r_t}\right)^{d-1}}} + \frac{(d-1)\delta^2}{2(d-3)} \left(\frac{r_h}{r_t}\right)^{d-1} \right.\nonumber\\
&& \left. \times \int_0^1 \frac{u^{2d-3}du}{\sqrt{1- u^{2(d-2)}}} \frac{\left(1-\left(\frac{r_h u}{r_t}\right)^{d-3} \right)}{\left(1-\left(\frac{r_h u}{r_t}\right)^{d-1} \right)^{3/2}} \right]. \nonumber \\
\end{eqnarray}
It is not possible to evaluate both the above integrals analytically in this existing form. We use the following identities
\begin{eqnarray}{\label{binomial}}
 \frac{1}{\sqrt{1-y}} = \sum_{n=0}^\infty \frac{\Gamma (n+1/2)}{\sqrt{\pi}\Gamma (n+1)}y^n~; \quad
 \frac{1}{(1-y)^{\frac{3}{2}}}=\sum_{n=0}^\infty \frac{2 \Gamma(n+3/2)}{\sqrt{\pi}\Gamma (n+1)}y^n
\end{eqnarray}
to express the above integrals in a convenient form which makes analytical solution possible. We rewrite eq.(\ref{3.65}) as 
\begin{eqnarray}{\label{lrt}}
 \frac{lr_t}{2} &=& \sum_{n=0}^{\infty} \frac{\Gamma(n+\frac{1}{2})}{\sqrt{\pi}\Gamma(n+1)} \left(\frac{r_h}{r_t}\right)^{(d-1)n} \int_0^1 \frac{u^{(d-1)n+d-2}du}{\sqrt{1-u^{2(d-2)}}}   \nonumber \\
&&+  \frac{(d-1)\delta^2}{2(d-3)} \left(\frac{r_h}{r_t}\right)^{d-1} \sum_{n=0}^{\infty} \frac{2\Gamma(n+\frac{3}{2})}{\sqrt{\pi}\Gamma(n+1)}  \left(\frac{r_h}{r_t}\right)^{(d-1)n}\int_0^1 \frac{u^{(d-1)n+2d-3}du}{\sqrt{1-u^{2(d-2)}}} \left(1-\left(\frac{r_h u}{r_t}\right)^{d-3} \right) \nonumber \\
\Rightarrow lr_t &= &\sum_{n=0}^{\infty} \frac{\Gamma(n+\frac{1}{2})}{(d-2)\Gamma(n+1)} \frac{\Gamma\left(\frac{(d-1)n+d-1}{2(d-2)}\right)}{\Gamma\left(\frac{(d-1)n+2d-3}{2(d-2)}\right)} \left(\frac{r_h}{r_t}\right)^{(d-1)n} + \frac{(d-1)\delta^2}{d-3} \sum_{n=0}^{\infty} \frac{\Gamma(n+\frac{3}{2})}{(d-2)\Gamma(n+1)}  \nonumber \\
&& \times \left(\frac{r_h}{r_t}\right)^{(d-1)n+d-1}  \left[ \frac{\Gamma\left(\frac{(d-1)n+2d-2}{2(d-2)}\right)}{\Gamma\left(\frac{(d-1)n+3d-4}{2(d-2)}\right)} - \left(\frac{r_h}{r_t}\right)^{d-3} \frac{\Gamma\left(\frac{(d-1)n+3d-5}{2(d-2)}\right)}{\Gamma\left(\frac{(d-1)n+4d-7}{2(d-2)}\right)} \right] ~.
\end{eqnarray}
Let us now look at the form of divergence for different terms of the above expression. For large value of $n$,
the first term behaves as $\sim \frac{1}{n}\left(\frac{r_h}{r_t}\right)^{n}$ and the second and third terms
behave in same way as $ \sim \left(\frac{r_h}{r_t}\right)^{n}$. Therefore the divergences 
of the second and third term cancels out each other. We isolate the divergent terms to get
\begin{eqnarray}
lr_t &=& \frac{\sqrt{\pi}}{d-2} \frac{\Gamma(\frac{d-1}{2(d-2)})}{\Gamma(\frac{2d-3}{2(d-2)})} + \sum_{n=1}^{\infty} \left[ \frac{\Gamma(n+\frac{1}{2})}{(d-2)\Gamma(n+1)}\frac{\Gamma(\frac{(d-1)n+d-1}{2(d-2)})}{\Gamma(\frac{(d-1)n+2d-3}{2(d-2)})} -\sqrt{\frac{2}{(d-2)(d-1)}} \frac{1}{n} \right] \left(\frac{r_h}{r_t}\right)^{(d-1)n} \nonumber \\
&+& \frac{(d-1)\delta^2}{(d-3)(d-2)} \sum_{n=0}^{\infty} \left[ \frac{\Gamma(n+\frac{3}{2})}{\Gamma(n+1)}\frac{\Gamma(\frac{(d-1)n+2d-2}{2(d-2)})}{\Gamma(\frac{(d-1)n+3d-4}{2(d-2)})} -\sqrt{\frac{2(d-2)}{(d-1)}}  \right] \left(\frac{r_h}{r_t}\right)^{(d-1)n+d-1} \nonumber \\
&-& \frac{(d-1)\delta^2}{(d-3)(d-2)} \sum_{n=0}^{\infty} \left[ \frac{\Gamma(n+\frac{3}{2})}{\Gamma(n+1)}\frac{\Gamma(\frac{(d-1)n+3d-5}{2(d-2)})}{\Gamma(\frac{(d-1)n+4d-7}{2(d-2)})} -\sqrt{\frac{2(d-2)}{(d-1)}}  \right] \left(\frac{r_h}{r_t}\right)^{(d-1)n+2d-4} \nonumber \\
&+& \frac{\delta^2}{d-3} \sqrt{\frac{2(d-1)}{d-2}} \frac{\left(1-\left(\frac{r_h}{r_t}\right)^{d-3}\right)}{\left(1-\left(\frac{r_h}{r_t}\right)^{d-1}\right)} \left(\frac{r_h}{r_t}\right)^{d-1} - \sqrt{\frac{2 }{(d-1)(d-2)}} \log\left(1- \left(\frac{r_h}{r_t}\right)^{d-1}\right) ~.
\end{eqnarray}
Now in the high temperature limit, $r_h$ takes large value. We can say that $r_h \sim r_t$. Hence we can write $r_t = (1+\epsilon) r_h$ where $\epsilon$ is a very small positive number. With this the above equation reads
\begin{eqnarray}
lr_h = -\sqrt{\frac{2}{(d-1)(d-2)}} \log((d-1)\epsilon)+ C_1 + \delta^2 C_2 + \mathcal{O}(\epsilon)
\label{eel1}
\end{eqnarray} 
where
\begin{eqnarray}
C_1 &=& \frac{\sqrt{\pi}}{d-2} \frac{\Gamma(\frac{d-1}{2(d-2)})}{\Gamma(\frac{2d-3}{2(d-2)})} + \sum_{n=1}^{\infty} \left[ \frac{\Gamma(n+\frac{1}{2})}{(d-2)\Gamma(n+1)}\frac{\Gamma(\frac{(d-1)n+d-1}{2(d-2)})}{\Gamma(\frac{(d-1)n+2d-3}{2(d-2)})} -\sqrt{\frac{2}{(d-2)(d-1)}} \frac{1}{n} \right] \nonumber \\
C_2 &=& \sqrt{\frac{2}{(d-2)(d-1)}} + \frac{(d-1)}{(d-3)(d-2)} \sum_{n=0}^{\infty} \left[ \frac{\Gamma(n+\frac{3}{2})}{\Gamma(n+1)}\frac{\Gamma(\frac{(d-1)n+2d-2}{2(d-2)})}{\Gamma(\frac{(d-1)n+3d-4}{2(d-2)})} -\sqrt{\frac{2(d-2)}{(d-1)}} \right] \nonumber \\
&& - \frac{(d-1)\delta^2}{(d-3)(d-2)} \sum_{n=0}^{\infty} \left[ \frac{\Gamma(n+\frac{3}{2})}{\Gamma(n+1)}\frac{\Gamma(\frac{(d-1)n+3d-5}{2(d-2)})}{\Gamma(\frac{(d-1)n+4d-7}{2(d-2)})} -\sqrt{\frac{2(d-2)}{(d-1)}}  \right] ~.
\end{eqnarray}
From the definition of $\delta$, we had earlier obtained 
\begin{eqnarray}
T= \frac{(d-1)r_h}{4\pi} (1-\delta^2).
\end{eqnarray}
Substituting the value of $r_h$ from the above equation in eq.(\ref{eel1}) and simplifying, we obtain
\begin{eqnarray}
\epsilon \approx \epsilon_{ent} e^{-\sqrt{\frac{d-2}{2(d-1)}} 4\pi Tl(1+\delta^2)}
\end{eqnarray}
where $\epsilon_{ent} = \frac{1}{d-1} e^{C_1 + C_2\delta^2}$. Now the surface area reads
\begin{eqnarray}
\mathcal{A} &=& 2 (Lr_t)^{d-3} \left[ \int_0^1 du \frac{u^{-d+2}}{\sqrt{1-u^{2(d-2)}}}\frac{1}{\sqrt{1- \left(\frac{r_h u}{r_t}\right)^{d-1}}} \right. \nonumber\\
&&\hspace{20mm}\left.  +\; \frac{(d-1)\delta^2}{2(d-3)} \left(\frac{r_h}{r_t}\right)^{d-1} 
 \int_0^1 du \frac{u\left(1- \left(\frac{r_h u}{r_t}\right)^{d-3}\right) }{\sqrt{1-u^{2(d-2)}}\left(1-\left(\frac{r_h u}{r_t}\right)^{d-1}\right)^{3/2}}\right] \nonumber \\
&=& 2 (Lr_t)^{d-3} \left[ \int_0^1 du \frac{u^{-d+2}}{\sqrt{1-u^{2(d-2)}}} \left(1+ \sum_{n=1}^{\infty} \frac{\Gamma(n+\frac{1}{2})}{\sqrt{\pi}\Gamma(n+1)} \left(\frac{r_h u}{r_t}\right)^{(d-1)n} \right) + \frac{(d-1)\delta^2}{2(d-3)} \left(\frac{r_h}{r_t}\right)^{d-1} \right. \nonumber \\
&& \left. ~~~~~~\times \int_0^1 du \frac{u\left(1- \left(\frac{r_h u}{r_t}\right)^{d-3}\right) }{\sqrt{1-u^{2(d-2)}}}\left( \sum_{n=0}^{\infty} \frac{2\Gamma(n+\frac{3}{2})}{\sqrt{\pi}\Gamma(n+1)} \left(\frac{r_h u}{r_t}\right)^{(d-1)n}\right)\right] ~. \nonumber\\
\end{eqnarray}
In the last line of the above equation we have used the relation \eqref{binomial} to evaluate 
the integrals analytically. The first integral corresponds to the area integral for pure AdS
spacetime. Only this integral is divergent near $u \rightarrow 0$. 
To remove this divergence, we introduce UV cut-off $\frac{1}{r_b}$ and add a counter term to obtain a finite value of surface area
\begin{eqnarray}
\mathcal{A}^{finite} &=& 2 (Lr_t)^{d-3} \left[ \int_{\frac{r_t}{r_b}}^1 du \frac{u^{-d+2}}{\sqrt{1-u^{2(d-2)}}} +  \sum_{n=1}^{\infty} \frac{\Gamma(n+\frac{1}{2})}{\sqrt{\pi}\Gamma(n+1)} \left(\frac{r_h}{r_t}\right)^{(d-1)n} \int_{0}^1 du \frac{u^{(d-1)n-d+2}}{\sqrt{1-u^{2(d-2)}}}   \right. \nonumber \\
&& \left. + \frac{(d-1)\delta^2}{2(d-3)} \sum_{n=0}^{\infty} \frac{2\Gamma(n+\frac{3}{2})}{\sqrt{\pi}\Gamma(n+1)} \left(\frac{r_h u}{r_t}\right)^{(d-1)n+d-1} \int_0^1 du \frac{u^{(d-1)n+1}}{\sqrt{1-u^{2(d-2)}}} \left(1- \left(\frac{r_h}{r_t}\right)^{d-3}\right) \right] \nonumber\\
&&-\frac{2(Lr_b)^{d-3}}{d-3} \nonumber\\
&=& (Lr_t)^{d-3} \left[ \frac{\sqrt{\pi}}{d-2}\frac{\Gamma(\frac{3-d}{2(d-2)})}{\Gamma(\frac{1}{2(d-2)})} +
\sum_{n=1}^{\infty} \frac{1}{d-2} \frac{\Gamma(n+1/2)\Gamma \left(\frac{n+1+d(n-1)}{2(d-2)} \right)}{\Gamma(n+1)\Gamma \left(\frac{n+1+nd}{2(d-2)}\right)}\left(\frac{r_h}{r_t}\right)^{(d-1)n}\right. \nonumber\\
&& \left.+ \frac{\delta^2 (d-1)}{(d-2)(d-3)} \sum_{n=0}^{\infty} \frac{\Gamma(n+3/2)\Gamma \left(\frac{2+n(d-1)}{2(d-2)}\right)}{\Gamma(n+1)\Gamma \left(\frac{d+n(d-1)}{2(d-2)}\right)}\left(\frac{r_h}{r_t}\right)^{(n+1)(d-1)} \right.\nonumber\\
&& \left. -\frac{\delta^2 (d-1)}{(d-2)(d-3)} \sum_{n=0}^{\infty} \frac{\Gamma(n+3/2)\Gamma \left(\frac{(d-1)(n+1)}{2(d-2)}\right)}{\Gamma(n+1)\Gamma \left(\frac{n(d+1)+2d-3}{2(d-2)}\right)} \left(\frac{r_h}{r_t}\right)^{n(d-1)+2(d-2)} \right] \nonumber\\
\end{eqnarray}
\begin{eqnarray}
&=& (Lr_t)^{d-3} \left[ \frac{\sqrt{\pi}}{d-2}\frac{\Gamma(\frac{3-d}{2(d-2)})}{\Gamma(\frac{1}{2(d-2)})}  \right. \nonumber\\
 && \left.+\sum_{n=1}^{\infty} \frac{1}{d-2}\left(1+\frac{d-2}{n+1+(n-1)(d-2)}\right) \frac{\Gamma(n+1/2)\Gamma \left(\frac{(n+1)(d-1)}{2(d-2)} \right)}{\Gamma(n+1)\Gamma \left(\frac{n+1+(n+2)(d-2)}{2(d-2)}\right)}\left(\frac{r_h}{r_t}\right)^{(d-1)n}\right. \nonumber\\
&& \left.+ \frac{\delta^2 (d-1)}{(d-2)(d-3)} \sum_{n=0}^{\infty}\left(1+\frac{d-2}{2+n(d-1)}\right) \frac{\Gamma(n+3/2)\Gamma \left(\frac{(n+2)(d-1)}{2(d-2)}\right)}{\Gamma(n+1)\Gamma \left(\frac{n+2+(n+3)(d-2)}{2(d-2)}\right)}\left(\frac{r_h}{r_t}\right)^{(n+1)(d-1)} \right.\nonumber\\
&& \left. -\frac{\delta^2 (d-1)}{(d-2)(d-3)} \sum_{n=0}^{\infty}\left(1+\frac{d-2}{n(d-1)+d-1}\right) \frac{\Gamma(n+3/2)\Gamma \left(\frac{n+1+(n+3)(d-2)}{2(d-2)}\right)}{\Gamma(n+1)\Gamma \left(\frac{n+1+(n+4)(d-2)}{2(d-2)}\right)} \left(\frac{r_h}{r_t}\right)^{n(d-1)+2(d-2)} \right] ~. \nonumber\\
\end{eqnarray}
Now we can use eq.(\ref{lrt}) to recast the surface area as
\begin{eqnarray}{\label{3.75}}
\mathcal{A}^{finite}&=& (Lr_t)^{d-3} \left[ \frac{\sqrt{\pi}}{3-d}\frac{\Gamma(\frac{d-1}{2(d-2)})}{\Gamma(\frac{2d-3}{2(d-2)})} + lr_t + \sum_{n=1}^{\infty} \frac{\Gamma(n+\frac{1}{2})\Gamma(\frac{(d-1)n+d-1}{2(d-2)}) \left(\frac{r_h}{r_t}\right)^{(d-1)n}}{\left( (d-1)n-(d-3)\right)\Gamma(n+1)\Gamma(\frac{(d-1)n+2d-3}{2(d-2)})} +\frac{(d-1)\delta^2}{d-3}  \right. \nonumber \\
&& \left. \times \sum_{n=0}^{\infty}  \frac{\Gamma(n+\frac{3}{2})}{\Gamma(n+1)} \left\{ \frac{\Gamma(\frac{(d-1)n+2(d-1)}{2(d-2)})\left(\frac{r_h}{r_t}\right)^{(d-1)n+d-1}}{\left( (d-1)n+2\right)\Gamma(\frac{(d-1)n+3d-4}{2(d-2)})} -\frac{\Gamma(\frac{(d-1)n+3d-5}{2(d-2)})\left(\frac{r_h}{r_t}\right)^{(d-1)n+2d-4}}{( (d-1)n+d-1)\Gamma(\frac{(d-1)n+4d-7}{2(d-2)})} \right\} \right] ~. \nonumber \\
\end{eqnarray}
The first summation term goes as $\sim \frac{1}{n^2}\left(\frac{r_h}{r_t}\right)^n$, where as
the second and third summation terms goes as  $\sim \frac{1}{n}\left(\frac{r_h}{r_t}\right)^n$ for large $n$. 
The leading contribution to the area $\mathcal{A}^{finite}$ comes from $r_h=r_t$. Those summation terms are not divergent in this limit. But as we know from \cite{Hubeny:2012ry} that the turning point of the extremal surface cannot penetrate the horizon radius $r_h$, we have to 
substitute $r_t=r_h(1+\epsilon)$ in eq.(\ref{3.75}) and expand binomially to get terms
that are higher order in $\epsilon$. Then it is easy to see that the summation terms are not convergent at the order $\mathcal{O}(\epsilon)$ and also at higher order. 
As in this case $r_h$ is very large, $\epsilon$ is likely to be a tiny quantity. 
So we consider contribution to surface area only upto order $\mathcal{O}(\epsilon)$.
We now use $r_t=r_h(1+\epsilon)$ and separate the divergent terms from the summations
to express eq.(\ref{3.75}) as
\begin{eqnarray}
 A^{finite}&=& L^{d-3} l r_h^{d-2} + L^{d-3}r_h^{d-3}(K_1+\delta^2 K_2)\nonumber \\
 &&+L^{d-3}r_h^{d-3}(K_3 \epsilon +\delta^2(K_4 \epsilon +K_5 \epsilon \log \epsilon))
\end{eqnarray}
where
\begin{eqnarray}
 K_1 &=& \frac{\sqrt{\pi}}{3-d} \frac{\Gamma \left(\frac{d-1}{2(d-2)}\right)}{\Gamma \left(\frac{2d-3}{2(d-2)}\right)}+ \frac{\sqrt{2(d-2)}}{(d-1)^{\frac{3}{2}}}\xi(2) \nonumber\\
 && +\sum_{n=1}^{\infty} \left(\frac{1}{((d-1)n+3-d)}\frac{\Gamma(n+1/2)\Gamma \left(\frac{(n+1)(d-1)}{2(d-2)}\right)}{\Gamma(n+1)\Gamma\left(\frac{(d-1)n+2d-3}{2(d-2)}\right)} -\frac{1}{n^2}\frac{\sqrt{2(d-2)}}{(d-1)^{\frac{3}{2}}}\right) \nonumber \\
 K_2 &=& \left(\frac{d-1}{d-3}\right)\left[\left(\frac{\Gamma \left(\frac{d-1}{d-2}\right)}{2\Gamma \left(\frac{3d-4}{2(d-2)}\right)} -\frac{\Gamma\left(\frac{3d-5}{2(d-2)}\right)}{(d-1)\Gamma \left(\frac{4d-7}{2(d-2)} \right)}\right)\Gamma(3/2) \right. \nonumber \\
 && \left. + \sum_{n=1}^{\infty} \left(\frac{1}{2+n(d-1)}\frac{\Gamma \left(n+\frac{3}{2}\right)\Gamma \left(\frac{(d-1)(n+2)}{2(d-2)}\right)}{\Gamma(n+1)\Gamma \left(\frac{(d-1)n+3d-4}{2(d-2)}\right)}- \frac{1}{n}\frac{\sqrt{2(d-2)}}{(d-1)^{\frac{3}{2}}}\right) \right. \nonumber \\
  && \left. + \sum_{n=1}^{\infty} \left(\frac{1}{(n+1)(d-1)}\frac{\Gamma \left(n+\frac{3}{2}\right)\Gamma \left(\frac{(d-1)n++3d-5}{2(d-2)}\right)}{\Gamma(n+1)\Gamma \left(\frac{(d-1)n+3d-7}{2(d-2)}\right)}- \frac{1}{n}\frac{\sqrt{2(d-2)}}{(d-1)^{\frac{3}{2}}}\right) \right. \nonumber \\
  K_3 &=& \sqrt{\frac{2(d-2)}{d-1}} \left(\log (d-1) -1\right) \nonumber \\
  K_4 &=& \left(\frac{d-1}{d-3}\right) \left[\frac{2(d-2)\Gamma \left(\frac{3d-5}{2(d-2)}\right)}{(d-1)\Gamma \left(\frac{4d-7}{2(d-2)}\right)} -\frac{(d-1)\Gamma\left(\frac{d-1}{d-2}\right)}{2 \Gamma\left(\frac{3d-4}{2(d-2)}\right)}\right]\Gamma\left(\frac{3}{2}\right) -\sqrt{\frac{2(d-2)}{d-1}}\log(d-1) \nonumber\\
  K_5 &=& -\sqrt{\frac{2(d-2)}{d-1}} ~.
\end{eqnarray}

\noindent Therefore the renormalized HEE in the small charge regime is given by 
\begin{eqnarray}
S_{A}^{finite} &=& L^{d-3}l S_{BH} + \frac{(Lr_h)^{d-3}}{4G_N^d}(K_1 +\delta^2 K_2) \nonumber \\
&& +\frac{L^{d-3}r_h^{d-3}}{4G_N^d}(K_3 \epsilon +\delta^2(K_4 \epsilon +K_5 \epsilon \log \epsilon))
\end{eqnarray} 
where $S_{BH} = \frac{r_h^{d-2}}{4G_N^d}~$.

\subsubsection{Large charge limit}
The HEE of a non-extremal $AdS$-RN black hole with high charge limit has been computed
in this section. The non-extremality condition (\ref{extremality}) sets the horizon radius ($r_h$) at large value and we have
$r_h l\gg 1$. Hence all the assumptions made in section (\ref{sel}) are also applicabe in this case.
Hence we can now Taylor expand eq.(\ref{lapseue}) around $u_0=\frac{r_t}{r_h}$ and neglect higher order terms to obtain
the lapse function as
\begin{eqnarray}
f(u) &=& f(u_0) + f^{\prime}(u_0) (u-u_0) + \mathcal{O}((u-u_0)^2) \nonumber \\
 &\approx& \left[(d-1)-\frac{(d-3)Q^2}{r_h^{2(d-2)}} \right]\left(1- \frac{u}{u_0}\right) ~.
\end{eqnarray}
Let us denote the prefactor infront of the above equation by $\sigma$, so that $\sigma =(d-1)-\frac{(d-3)Q^2}{r_h^{2(d-2)}}$.
This factor also had appeared in the expression for black hole temperature in eq.(\ref{bhtemp}).
In the low temperature limit we have $\sigma \rightarrow 0$ and in high temperature limit we have  $\sigma \rightarrow (d-1)$.
Now the length of the subsystem reads
\begin{eqnarray}{\label{3.80}}
l &=& \frac{2}{r_t\sqrt{\sigma}} \int_0^1 \frac{u^{d-2}du}{\sqrt{1-u^{2(d-2)}}} \frac{1}{\sqrt{1- \frac{r_h u}{r_t}}} \nonumber \\
\Rightarrow lr_t &=& \frac{2}{\sqrt{\sigma}} \sum_{n=0}^{\infty} \frac{\Gamma(n+\frac{1}{2})}{\sqrt{\pi}\Gamma(n+1)} \left(\frac{r_h}{r_t}\right)^n \int_0^1 \frac{u^{n+d-2} du}{\sqrt{1-u^{2(d-2)}}} \nonumber \\
&=& \frac{1}{(d-2)\sqrt{\sigma}} \sum_{n=0}^{\infty} \frac{\Gamma(n+\frac{1}{2})}{\sqrt{\pi}\Gamma(n+1)} \frac{\Gamma(\frac{n+d-1}{2(d-2)})}{\Gamma(\frac{n+2d-3}{2(d-2)})} \left(\frac{r_h}{r_t}\right)^n
\end{eqnarray}  
where we have used eq.(\ref{binomial}). 
For large value of $n$, this expression is divergent. Using gamma function properties and Stirling formula, 
one can see that for large value of $n$, the above summation goes 
as $\frac{\sqrt{2}}{(d-2)\;n} \left(\frac{r_h}{r_t}\right)^n $. Therefore it is divergent as $r_t \rightarrow r_h$.
To get a finite value we isolate the divergent terms to get  
\begin{eqnarray}
lr_t &=& \frac{\sqrt{\pi}}{(d-2)\sqrt{\sigma}} \frac{\Gamma(\frac{d-1}{2(d-2)})}{\Gamma(\frac{2d-3}{2(d-2)})} + \frac{1}{\sqrt{\sigma}} \sum_{n=1}^{\infty} \left( \frac{\Gamma(n+\frac{1}{2})}{(d-2)\Gamma(n+1)} \frac{\Gamma(\frac{n+d-1}{2(d-2)})}{\Gamma(\frac{n+2d-3}{2(d-2)})} - \sqrt{\frac{2}{d-2}}\frac{1}{n} \right) \left(\frac{r_h}{r_t}\right)^n \nonumber \\
&&- \sqrt{\frac{2}{(d-2)\sigma}} \log\left(1- \frac{r_h}{r_t}\right)  ~.
\end{eqnarray}
Now we substitute $r_t=r_h (1+\epsilon)$ and expand around $\epsilon$ to finally obtain
\begin{eqnarray}
\sqrt{\sigma }lr_h = - \sqrt{\frac{2}{d-2}} \log(\epsilon) + D_1 + \mathcal{O}(\epsilon) 
\label{eel4}
\end{eqnarray}
where 
\begin{eqnarray}
D_1 = \frac{\sqrt{\pi}}{(d-2)} \frac{\Gamma(\frac{d-1}{2(d-2)})}{\Gamma(\frac{2d-3}{2(d-2)})} +  \sum_{n=1}^{\infty} \left\{ \frac{\Gamma(n+\frac{1}{2})}{(d-2)\Gamma(n+1)} \frac{\Gamma(\frac{n+d-1}{2(d-2)})}{\Gamma(\frac{n+2d-3}{2(d-2)})} - \sqrt{\frac{2}{d-2}}\frac{1}{n} \right\} ~.
\end{eqnarray}
From the above equation, one can find
\begin{eqnarray}
\epsilon = \epsilon_{ent} e^{-\frac{(d-2)\sigma}{2}lr_h} 
\end{eqnarray}
where $\epsilon_{ent} = e^{\frac{d-2}{2}D_1}$. 
Now we shall compute the extremal surface area which reads
\begin{equation}
 \mathcal{A} = \frac{2(Lr_t)^{d-3}}{\sqrt{\sigma}} \int_0^1 du\frac{1}{u^{d-2}\sqrt{1-u^{2(d-2)}}}\frac{1}{\sqrt{\sigma \left(1-\frac{r_h}{r_t}u\right)}} ~.
\end{equation}
Since we want an analytical expression of area we use eq.(\ref{binomial}) to rewrite the area integral as
\begin{eqnarray}
\mathcal{A} = \frac{2(Lr_t)^{d-3}}{\sqrt{\sigma}} \sum_{n=0}^{\infty} \frac{\Gamma(n+\frac{1}{2})}{\sqrt{\pi}\Gamma(n+1)} \left(\frac{r_h}{r_t}\right)^n \int_{0}^{1} \frac{u^{n-d+2}du}{\sqrt{1-u^{2(d-2)}}} ~.
\end{eqnarray}
This integral seems to be divergent for $n < (d-2)$. The divergence occurs when $u \rightarrow 0$.
So we can put an UV cutoff $r_t/r_b$ in the integrals to regularize it from $n=0$ to $n=d-3$. We use
the same procedure as used in section (\ref{sel}) to obtain the finite part of area integrals as
\begin{eqnarray}
 \mathcal{A}^{finite}_0 &=&-\frac{2\sqrt{\pi} (Lr_t)^{d-3}}{(d-3)\sqrt{\sigma}} \frac{\Gamma \left( \frac{d-1}{2(d-2)}  \right)}{\Gamma \left(\frac{1}{2(d-2)}   \right)} \nonumber\\
  \mathcal{A}^{finite}_m &=&  \frac{2(Lr_t)^{d-3}}{\sqrt{\sigma}}\frac{\Gamma(n+\frac{1}{2})}{\sqrt{\pi}\Gamma(n+1)} \left(\frac{r_h}{r_t}\right)^m 
	\left[\frac{1}{d-m-3}+ \frac{\sqrt{\pi}}{2(d-2)}\frac{\Gamma (\frac{m-d+3}{2(d-2)})}{\Gamma (\frac{m+1}{2(d-2)})} \right];\mbox{(m=1,2,...,$d-4$)}\nonumber\\
  \mathcal{A}^{finite}_{d-3}&=& \frac{2(Lr_t)^{d-3}\Gamma \left(\frac{2d-5}{2}\right)}{\sqrt{\pi \sigma \Gamma(d-2)}} \left(\frac{r_h}{r_t}\right)^{d-3}\frac{\log 4}{2(d-2)} ~.
	\end{eqnarray}
For $n \geq (d-2)$, we get
\begin{eqnarray}
\mathcal{A}_{n\geq (d-2)} &=& \frac{2(Lr_t)^{d-3}}{\sqrt{\sigma}} \sum_{n=(d-2)}^{\infty} \frac{\Gamma(n+\frac{1}{2})}{\sqrt{\pi}\Gamma(n+1)} \left(\frac{r_h}{r_t}\right)^n \int_{0}^{1} \frac{u^{n-d+2}du}{\sqrt{1-u^{2(d-2)}}} \nonumber \\
&=& \frac{2(Lr_t)^{d-3}}{\sqrt{\sigma}} \sum_{n=(d-2)}^{\infty} \frac{1}{2(d-2)}\frac{\Gamma(n+\frac{1}{2})}{\Gamma(n+1)}  \frac{\Gamma(\frac{n-d+3}{2(d-2)})}{\Gamma(\frac{n+1}{2(d-2)})} \left(\frac{r_h}{r_t}\right)^n \nonumber \\
&=& \frac{(Lr_t)^{d-3}}{\sqrt{\sigma}}  \sum_{n=(d-2)}^{\infty}  \left( \frac{1}{d-2}+ \frac{1}{n-d+3} \right) \frac{\Gamma(n+\frac{1}{2})}{\Gamma(n+1)} \frac{\Gamma(\frac{n+d-1}{2(d-2)})}{\Gamma(\frac{n+2d-3}{2(d-2)})} \left(\frac{r_h}{r_t}\right)^n .
\end{eqnarray}
Using eq.(\ref{3.80}) we can recast the above expression  in terms of the subsystem length as below
\begin{eqnarray}{\label{3.89}}
 \mathcal{A}_{n\geq (d-2)}^{finite} &=& \frac{(Lr_t)^{d-3}}{\sqrt{\sigma}} \left[ \sqrt{\sigma}\; l r_t -\sum_{m=0}^{d-3} \frac{1}{d-2} \frac{\Gamma(m+\frac{1}{2})}{\Gamma(m+1)} \frac{\Gamma(\frac{m+d-1}{2(d-2)})}{\Gamma(\frac{m+2d-3}{2(d-2)})} \left(\frac{r_h}{r_t}\right)^m \nonumber \right. \\
&& \left. + \sum_{n=d-2}^{\infty} \frac{1}{n-d+3} \frac{\Gamma(n+\frac{1}{2})}{\Gamma(n+1)} \frac{\Gamma(\frac{n+d-1}{2(d-2)})}{\Gamma(\frac{n+2d-3}{2(d-2)})} \left(\frac{r_h}{r_t}\right)^n   \right]~. \nonumber \\
\end{eqnarray}
For large value of $n$, the third term goes as $\frac{\sqrt{2(d-2)}}{n^2} \left(\frac{r_h}{r_t}\right)^n$.
Hence this term is finite as $r_t \rightarrow r_h$. Hence we can get a finite leading order contribution to
this term. But as $r_t=r_h(1+\epsilon)$, one can see that the third term in eq.(\ref{3.89}) is divergent at first order in $\epsilon$.
So we separate the divergent terms to rewrite eq.(\ref{3.89}) as
\begin{eqnarray}
\mathcal{A}_{n\geq (d-2)}^{finite} &=& \frac{(Lr_t)^{d-3}}{\sqrt{\sigma}} \left[ \sqrt{\sigma} lr_t -\sum_{m=0}^{d-3} \frac{1}{d-2} \frac{\Gamma(m+\frac{1}{2})}{\Gamma(m+1)} \frac{\Gamma(\frac{m+d-1}{2(d-2)})}{\Gamma(\frac{m+2d-3}{2(d-2)})} \left(\frac{r_h}{r_t}\right)^m \nonumber \right. \\
&& \left. + \sum_{n=d-2}^{\infty} \left\{ \frac{1}{n-d+3} \frac{\Gamma(n+\frac{1}{2})}{\Gamma(n+1)} \frac{\Gamma(\frac{n+d-1}{2(d-2)})}{\Gamma(\frac{n+2d-3}{2(d-2)})} -  \frac{\sqrt{2(d-2)}}{n^2}\right\} \left(\frac{r_h}{r_t}\right)^n \right. \nonumber \\
&& \left. + \sum_{n=d-2}^{\infty} \frac{\sqrt{2(d-2)}}{n^2} \left(\frac{r_h}{r_t}\right)^n  \right]~.
\end{eqnarray}
Now we can write 
\begin{eqnarray}
\sum_{n=d-2}^{\infty} \frac{\sqrt{2(d-2)}}{n^2} \left(\frac{r_h}{r_t}\right)^n = \sqrt{2(d-2)} \left[ Li_{2} \left[\frac{r_h}{r_t}\right]  - \sum_{m=1}^{d-3} \frac{1}{m^2} \left(\frac{r_h}{r_t}\right)^m \right].
\end{eqnarray}
Using this, the total surface area reads
\begin{eqnarray}
\mathcal{A}^{finite} &=& 
 \frac{(Lr_t)^{d-3}}{\sqrt{\sigma}} \left[ \sqrt{\sigma} lr_t -\frac{2\sqrt{\pi}(d-2)}{d-3}\frac{\Gamma \left(\frac{d-1}{2(d-2)}\right)}{\Gamma \left(\frac{1}{2(d-2)}\right)} \right.\nonumber \\ 
&&+ \left. \sum_{n=1}^{d-4} \frac{\Gamma(n+\frac{1}{2})}{\sqrt{\pi}\Gamma(n+1)} \left(\frac{\sqrt{\pi}}{n+1} \frac{\Gamma(\frac{n-d+3}{2(d-2)})}{\Gamma(\frac{n+1}{2(d-2)})} + \frac{2}{[d-(n+3)]}\right) \left(\frac{r_h}{r_t}\right)^n \right. \nonumber \\
 &&+ \left.\frac{\Gamma \left(\frac{2d-5}{2}\right)}{(d-2)\sqrt{\pi}\Gamma(d-2)}\left(\log 4-\Gamma(3/2)\right)\left(\frac{r_h}{r_t}\right)^{d-3}\right. \nonumber \\ 
 && \left. +\sqrt{2(d-2)} \left( Li_{2} \left[\frac{r_h}{r_t}\right] - \sum_{n=1}^{d-3} \frac{1}{n^2} \left(\frac{r_h}{r_t}\right)^n \right)\right. \nonumber\\
 && \left.   + \sum_{n=d-2}^{\infty} \left( \frac{1}{n-d+3}\frac{\Gamma(n+\frac{1}{2})}{\Gamma(n+1)}\frac{\Gamma(\frac{n+d-1}{2(d-2)})}{\Gamma(\frac{n+2d-3}{2(d-2)})} - \frac{\sqrt{2(d-2)}}{n^2}\right) \left(\frac{r_h}{r_t}\right)^n \right] ~. \nonumber \\
\label{ee80}
\end{eqnarray}
Now we substitute $r_t = r_h(1+\epsilon)$ in eq.(\ref{ee80}) to know the sub-leading term upto order $\epsilon$.
After simplification we finally obtain the finite part of the area of extremal surface to be 
\begin{eqnarray}
\mathcal{A}^{finite} = L^{d-3} l r_h^{d-2} + \frac{(Lr_h)^{d-3}}{\sqrt{\sigma}} \left\{ K_1^{'}   +K_2^{'} \epsilon+ \mathcal{O}(\epsilon^2) \right\}
\end{eqnarray} 
where 
\begin{eqnarray}
K_1^{'} &=&   -2\sqrt{\pi}\frac{d-2}{d-3}\frac{\Gamma(\frac{d-1}{2(d-2)})}{\Gamma(\frac{1}{2(d-2)})} + \sqrt{\frac{d-2}{2}}\xi(2) -\sqrt{\frac{d}{2}}\sum_{n=1}^{d-3}\frac{1}{n^2}  \nonumber \\
&&   + \sum_{n=1}^{d-4} \frac{\Gamma\left(n+\frac{1}{2}\right)}{\sqrt{\pi}\Gamma(n+1)} \left[  \frac{\sqrt{\pi}}{n+1}  \frac{\Gamma(\frac{n-d+3}{2(d-2)})}{\Gamma(\frac{n+1}{2(d-2)})} + \frac{2}{d-(n+3)}\right] \nonumber\\
&& + \frac{\Gamma \left(\frac{2d-5}{2}\right)}{\sqrt{\pi}(d-2)\Gamma(d-2)}\left(\log 4 -\frac{\sqrt{\pi}}{\Gamma(3/2)}\right) \nonumber \\
&& + \sum_{n=d-2}^{\infty} \left[\frac{1}{n-d+3} \frac{\Gamma(n+\frac{1}{2})}{\Gamma(n+1)} \frac{\Gamma(\frac{n+d-1}{2(d-2)})}{\Gamma(\frac{n+2d-3}{2(d-2)})} -\frac{\sqrt{2(d-2)}}{n^2} \right]  \nonumber \\
K_2^{'}&=&\frac{\Gamma \left(\frac{2d-7}{2}\right)}{\sqrt{\pi}\Gamma (d-3)}\left(2+\frac{\sqrt{\pi}\Gamma\left(\frac{-1}{2(d-2)}\right)}{(d-3)\Gamma \left(\frac{d-3}{2(d-2)}\right)}\right) -\sqrt{2(d-2)} ~.
\end{eqnarray}
Therefore the renormalized holographic entanglement entropy is given by (in the large charge regime for non-extremal black hole)
\begin{eqnarray}
S_{A}^{finite} = L^{d-3}l S_{BH}^{ext} + \frac{(Lr_h)^{d-3}}{4G_N\sqrt{\sigma}} \left\{ K_1^{'}   +K_2^{'} \epsilon+ \mathcal{O}(\epsilon^2) \right\}
\end{eqnarray} 
where $S_{BH}^{ext} = \frac{r_h^{d-2}}{4G_N^d}~$.



\section{Entanglement thermodynamics}
In this section, we investigate the entanglement thermodynamics in the small charge limit. 
To carry out this study, one has to note from the $AdS$/CFT duality that in the presence of a RN black hole, an extremal black hole has to be considered as the dual to the ground state (zero temperature state) of the boundary field theory. Further, a non-extremal black hole has to be considered as the dual to the excited state (finite temperature state) of the boundary field theory. The difference in the entanglement entropies of these two states would lead to the first law of entanglement thermodynamics. Hence we obtain 
\begin{eqnarray}
\Delta S_A = \frac{\Delta E_A}{T_{ent}}
\label{eet1}
\end{eqnarray}
where 
\begin{eqnarray}
\Delta S_A &=& S_A- S_A^{ext} = k L^{d-3} l^2 (M - M^{ext}) \\
\Delta E_A &=& \int_{A} dx_1 dx_2...dx_{d-3} T_{tt}^{temp\neq 0} -\int_{A} dx_{1} dx_2...dx_{d-3} T_{tt}^{temp= 0} \nonumber \\
&=& \frac{d-2}{16\pi G_N^d} L^{d-3} l (M-M^{ext}) ~.
\end{eqnarray}
Substituting these two relation in eq.(\ref{eet1}), the entanglement temperature reads
\begin{eqnarray}
T_{ent} = \frac{2(d-2)^2}{\sqrt{\pi}} \left(\frac{\Gamma(\frac{d-1}{2(d-2)})}{\Gamma(\frac{1}{2(d-2)})}\right)^2 \left[\frac{1}{\frac{\Gamma(\frac{1}{d-2})}{\Gamma(\frac{d}{2(d-2)})}-\frac{\Gamma(\frac{d-1}{2(d-2)})}{\Gamma(\frac{1}{2(d-2)})}}\right] ~.
\end{eqnarray}
\section{Conclusion}
We have explicitly investigated the entanglement thermodynamics for $d$-dimensional charged black hole by studying the holographic entanglement entropy in different cases. We computed the 
holographic entanglement entropy in extremal and non-extremal cases in two different regimes, namely, the small charge limit and the large charge limit. For non-extremal black hole, there are two limiting cases, namely, the low temperature limit
and the high temperature limit. We have calculated the holographic entanglement entropy for all these cases. It is observed that the holographic entanglement entropy for small charged 
extremal black hole is the same as the holographic entanglement entropy in the low temperature regime for small charged non-extremal black hole.
We have then found the first law of entanglement thermodynamics for boundary field theory in the low temperature regime in arbitrary dimensions for the small charge limit.
From this we have calculated the entanglement temperature for this system.

\section*{Acknowledgment}
DG would like to thank to DST-INSPIRE for financial support. S.~Gangopadhyay acknowledges the support by DST SERB under Start Up Research
Grant (Young Scientist), File No. YSS/2014/000180. He also acknowledges the support of IUCAA, Pune for the Visiting Associateship programme.

\end{document}